\documentstyle[11pt,aaspp4]{article}

\def\be{\begin{equation}}
\def\ee{\end{equation}}

\def\etal{{et al.~}}

\def\gs{\mathrel{\raise1.16pt\hbox{$>$}\kern-7.0pt
\lower3.06pt\hbox{{$\scriptstyle \sim$}}}}
\def\ls{\mathrel{\raise1.16pt\hbox{$<$}\kern-7.0pt
\lower3.06pt\hbox{{$\scriptstyle \sim$}}}}
\def\gtsima{$\; \buildrel > \over \sim \;$}
\def\ltsima{$\; \buildrel < \over \sim \;$}
\def\prosima{$\; \buildrel \propto \over \sim \;$}
\def\gsim{\lower.5ex\hbox{\gtsima}}
\def\lsim{\lower.5ex\hbox{\ltsima}}
\def\simgt{\lower.5ex\hbox{\gtsima}}
\def\simlt{\lower.5ex\hbox{\ltsima}}
\def\simpr{\lower.5ex\hbox{\prosima}}
\def\la{\lsim}
\def\ga{\gsim}

\def\pp{\noindent\parshape 2 0truecm 17truecm 2truecm 15truecm}
\def\rf#1;#2;#3;#4 {\par\pp#1, #2, #3, #4. \par}

\def\pr{\ref@jnl{Phys.Rev}}

\def\ie{{\frenchspacing\it i.e. }}

\def\href#1;#2 {{\bf #1} : {\em #2}}

\def\beq#1{\begin{equation}\label{#1}}
\def\eeq{\end{equation}}
\def\beqa#1{\begin{eqnarray}\label{#1}}
\def\eeqa{\end{eqnarray}}

\def\H2p{H$_2^+$ }

\def\mH2p{H_2^+}

\begin{document}

\title{Expected Number and Flux Distribution of Gamma-Ray-Burst 
Afterglows with High Redshifts}

\author{Benedetta Ciardi \altaffilmark{1,2} and Abraham Loeb \altaffilmark{2}}

\altaffiltext{1}{Universit\`a degli Studi di Firenze, Dipartimento di
Astronomia, L.go E. Fermi 5, Firenze, Italy}
\altaffiltext{2}{Harvard-Smithsonian Center for Astrophysics, 60 Garden Street,
Cambridge, MA 02138}

\begin{abstract}

If Gamma-Ray-Bursts (GRBs) occur at high redshifts, then their bright
afterglow emission can be used to probe the ionization and metal enrichment
histories of the intervening intergalactic medium during the epoch of
reionization.  In contrast to other sources, such as galaxies or quasars,
which fade rapidly with increasing redshift, the observed infrared flux
from a GRB afterglow at a fixed observed age is only a weak function of its
redshift.  This results from a combination of the spectral slope of GRB
afterglows and the time-stretching of their evolution in the observer's
frame.  Assuming that the GRB rate is proportional to the star formation
rate and that the characteristic energy output of GRBs is $\sim
10^{52}~{\rm ergs}$, we predict that there are always $\sim 15$ GRBs from
redshifts $z\ga 5$ across the sky which are brighter than $\sim 100$ nJy at
an observed wavelength of $\sim 2\mu$m.  The infrared spectrum of these
sources could be taken with the future {\it Next Generation Space
Telescope}, as a follow-up on their early X-ray localization with the {\it
Swift} satellite.

\end{abstract}
\keywords {gamma rays: bursts --- ISM}

\section{Introduction}

The past decade has been marked by major observational breakthroughs
concerning the properties of the Gamma Ray Burst (GRB) sources.  The Burst
and Transient Source Experiment (BATSE) on board the Compton Gamma Ray
Observatory (Meegan \etal 1992) showed that the GRB population is
distributed isotropically across the sky, and that there is a deficiency of
faint GRBs relative to a Euclidean distribution.  These were the first
observational clues indicating a cosmological distance scale for GRBs.  The
localization of GRBs by X-ray observations with the BeppoSAX satellite
(Costa \etal 1997) allowed the detection of afterglow emission at optical
(e.g., van Paradijs \etal 1997) and radio (e.g., Frail \etal 1997; Frail,
Waxman \& Kulkarni 1999) wavelengths up to more than a year following the
events (Fruchter et al. 1999; Frail, Waxman \& Kulkarni 1999).  The
afterglow emission is characterized by a broken power-law spectrum with a
peak frequency that declines with time.  The radiation can be well modeled
as synchrotron emission from a decelerating blast wave, created by the GRB
explosion in an ambient medium, plausibly the interstellar medium of
galaxies (Waxman 1997; Wijers \& Galama 1999; M\'esz\'aros 1999; but see
also Chevalier \& Li 1999).  The detection of spectral features, such as
metal absorption lines (Metzger \etal 1997), in some optical afterglows
allowed an unambiguous identification of the cosmological distance
scale. By now, the redshift of almost a dozen GRBs has been identified
either through the detection of absorption features in the afterglow
spectra or of emission lines from host galaxies (see Fig. 3 in Kulkarni et
al. 2000).

The central engine of GRBs is still unknown. Since the inferred energy
release is comparable to or higher than that in supernovae, most popular
models relate GRBs to stellar remnants, such as neutron stars or black
holes (see, e.g., Eichler \etal 1989; Paczy\'{n}ski 1991; Usov 1992;
Mochkovitch \etal 1993; Paczy\'{n}ski 1998; MacFadyen \& Woosley
1999). Recently it has been claimed that the late evolution of some rapidly
declining optical afterglows shows a component which is possibly associated
with supernova emission (e.g., Bloom \etal 1999; Reichart 1999). If the
supernova association will be confirmed by detailed spectra of future
afterglows, the GRB phenomenon will be linked to the terminal evolution of
massive stars.

Any association of GRBs with the formation of single compact stars implies
that the GRB rate should trace the star formation history of the universe
(Totani 1997; Sahu \etal 1997; Wijers \etal 1998; but see Krumholz,
Thorsett \& Harrison 1998).  Owing to their high brightness, GRB afterglows
might be detected out to exceedingly high redshifts. Similarly to quasars,
the broad band emission of GRB afterglows can be used to probe the
absorption properties of the intergalactic medium (IGM) out to the epoch
when it was reionized at a redshift $z\sim 10$ (Miralda-Escud\'e \& Rees
1998; Loeb 1999).  Lamb \& Reichart (1999) have extrapolated the observed
gamma-ray and afterglow spectra of known GRBs to high redshifts and
emphasized the important role that their detection might play in probing
the IGM.  Simple scaling of the long-wavelength spectra and temporal
evolution of afterglows with redshift implies that at a fixed time lag
after the GRB in the observer's frame, there is only mild change in the
{\it observed} flux at infrared or radio wavelengths with increasing
redshift. This results in part from the fact that afterglows are brighter
at earlier times, and that a given observed time refers to an earlier
intrinsic time in the source frame as the source redshift increases. The
mild dependence of the long-wavelength flux on redshift is in contrast with
other high-redshift sources such as galaxies or quasars, which fade rapidly
with increasing redshift (Haiman \& Loeb 1998; 1999).  The ``apparent
brightening'' of GRB afterglows with redshift could be further enhanced by
the expected increase in the mean density of the interstellar medium of
galaxies at increasing redshifts (Wood \& Loeb 1999).

It therefore appears natural to use GRBs as an important tool in
probing the high-redshift universe and its star formation history (Blain \&
Natarajan 1999).  Since GRBs are rare, all-sky searches for their early
$\gamma$-ray emission are needed before follow-up observations at much
longer wavelength are conducted.

In this paper we model the emission from GRB afterglows and follow their
number counts at different wavelengths as a function of redshift. For
simplicity, we assume that the ambient gas surrounding GRB sources is the
interstellar medium of their host galaxies.  In \S 2 we review our model
for both the relativistic and non-relativistic stages of the afterglow
emission, as well as a simple prescription for the redshift evolution of
the interstellar medium of galaxies.  The numerical results and their
dependence on model assumptions are discussed in \S \ref{res}. Finally, \S
\ref{disc} summarizes our main conclusions and describes the implications
of our results.

\section{Afterglow Emission}\label{aftem}

\subsection{Relativistic Regime}

GRB afterglows can be modeled as synchrotron emission from a decelerating
relativistic blast wave, created by the GRB explosion in an external
medium. For a point explosion in a uniform medium, the shock structure in
the highly-relativistic regime is described by the Blandford \& McKee
(1976) self-similar solution. The synchrotron emission originates from a
power-law population of shock-accelerated electrons in a strong magnetic
field (Waxman 1997; M\'esz\'aros, Rees, \& Wijers 1998; Sari, Piran \&
Narayan 1998), with both the electrons and the magnetic field having close
to equipartition energy densities (see Medvedev \& Loeb 1999 for the
possible origin of the magnetic field).  For simplicity, we focus our
discussion on a spherically-symmetric explosion in a uniform ambient
medium, assumed to be the interstellar medium of the host galaxy, although
at least some GRB blast waves might be expanding into the stellar wind of
the progenitor star (Dai \& Lu 1998; M\'esz\'aros, Rees \& Wijers 1998;
Chevalier \& Li 1999). The dependence of our results on the ambient medium
density will be discussed in \S~\ref{pardep}.
We assume that the shock-accelerated electrons have a power-law
distribution of Lorentz factor, $\gamma_e$, with a minimum Lorentz factor
$\gamma_m$ (see, e.g., Sari, Piran \& Narayan 1998). We also define
$\gamma_c$ as the threshold Lorentz factor below which electrons do not
lose a significant fraction of their energy to radiation. In the {\it fast
cooling} regime, when $\gamma_m>\gamma_c$, all of the electrons cool
rapidly down to a Lorentz factor $\sim \gamma_c$ and the flux observed at
the frequency $\nu$ is given by,
\begin{equation}
F_\nu=F_{\nu_m} \left\{
\begin{array}{ll}
(\nu/\nu_c)^{1/3} & \nu_c>\nu,\\
(\nu/\nu_c)^{-1/2} & \nu_m>\nu \ge \nu_c,\\
(\nu_m/\nu_c)^{-1/2}(\nu/\nu_m)^{-p/2} & \nu \ge \nu_m,\\
\end{array}\right.
\label{fnuf}
\end{equation}
where $F_{\nu_m}$ is the observed peak flux, $\nu_c= \nu(\gamma_c)$ and
$\nu_m=\nu(\gamma_m)$ are the characteristic synchrotron frequency
calculated at $\gamma_c$ and $\gamma_m$ respectively and $p$ is the
power-law index of the electron energy distribution. A typical value of
$p\approx2.5$ often fits both the GRB and the afterglow observations (Sari,
Piran \& Narayan 1998; Kumar \& Piran 1999).

When $\gamma_c>\gamma_m$, only electrons with $\gamma_e>\gamma_c$ cool
efficiently. In this {\it slow cooling} regime the observed flux is,
\begin{equation}
F_\nu=F_{\nu_m} \left\{
\begin{array}{ll}
(\nu/\nu_m)^{1/3} & \nu_m>\nu,\\
(\nu/\nu_m)^{-(p-1)/2} & \nu_c>\nu \ge \nu_m,\\
(\nu_c/\nu_m)^{-(p-1)/2}(\nu/\nu_c)^{-p/2} & \nu \ge \nu_c.\\
\end{array}\right.
\label{fnus}
\end{equation}
Typically, synchrotron self-absorption results in an additional break, at a
frequency $\la 5$GHz. We will not consider the low-frequency regime below
this break in our discussion.  Assuming a fully adiabatic shock (Sari,
Piran \& Narayan 1998),
\begin{equation}
\nu_c=2.7 \times 10^{12} \epsilon_B^{-3/2} E_{52}^{-1/2} n_1^{-1}
t_d^{-1/2} (1+z)^{-1/2} \; {\rm Hz},
\label{nuc}
\end{equation}
\begin{equation}
\nu_m=5.7 \times 10^{14} \epsilon_B^{1/2} \epsilon_e^{2}
E_{52}^{1/2} t_d^{-3/2} (1+z)^{1/2} \; {\rm Hz},
\label{num}
\end{equation}
\begin{equation}
F_{\nu_m}=1.1 \times 10^{5} \epsilon_B^{1/2} E_{52} n_1^{1/2}
d_{28}^{-2} (1+z) \; {\rm \mu Jy}.
\label{fmax}
\end{equation}
The redshift dependence results from the fact that the radiation emitted by
a source at a redshift $z$ at the frequency $\nu_s$ over a time $\Delta
t_s$, will be observed at $z=0$ at a frequency $\nu_o=\nu_s/(1+z)$ over a
time $\Delta t_o=(1+z) \Delta t_s$.  Here $\epsilon_B$ and $\epsilon_e$ are
the fraction of the shock energy that is converted to magnetic fields and
accelerated electrons, respectively, for which we adopt the values
$\epsilon_B=0.1$ and $\epsilon_e=0.2$ (see, e.g., Waxman 1997); $E=E_{52}
10^{52}$ erg is the energy of the spherical shock; $n =n_1$ 1 cm$^{-3}$ is
the mean number density of the ambient gas; $t=t_d$ 1 day is the time lag
since the GRB trigger, as measured in the observer frame, and $d_L=d_{28}
10^{28}$ cm is the cosmology-dependent luminosity distance. For a flat
universe ($\Omega_0=\Omega_m+\Omega_\Lambda=1$), the luminosity distance
can be written as:
\begin{equation}
d_L=(1+z) \int_0^z (1+z^\prime) \left\vert \frac{dt}{dz^\prime}
\right\vert dz^\prime,
\label{dl}
\end{equation}
where 
\begin{equation}
\left(\frac{dt}{dz}\right)^{-1}=-H_0 (1+z) \sqrt{(1+\Omega_mz)(1+z)^2-
\Omega_\Lambda(2z+z^2)}.
\label{dtdz}
\end{equation}
$H_0=100 h$ Km s$^{-1}$ Mpc$^{-1}$ is the current Hubble constant.
Throughout the paper we adopt a flat cosmology with $h=0.65$, density
parameters $\Omega_m=0.35$ and $\Omega_\Lambda=0.65$, and a Cold Dark
Matter power spectrum of density fluctuations with power-law index
$n=0.96$, an amplitude $\sigma_8=0.87$ and baryon density parameter
$\Omega_b=0.04$ (Bahcall \etal 1999).  The explosion energy, $E$, is highly
uncertain. Based on X-ray afterglow data, Freedman \& Waxman (1999)
inferred explosion energies in the range $10^{51.5}-10^{53.5}~{\rm
ergs}(\Delta \Omega/4\pi)$, where $\Delta \Omega$ is the solid angle into
which the energy is channeled. If one assumes that GRB explosions are
isotropic, then in some events the radiated energy is estimated to be in
excess of $10^{53}$ ergs, reaching a value of $3.4 \times 10^{54}$ erg for
GRB 990123 (see, e.g., Kulkarni \etal 1999).  It has been suggested that
these values are reduced by orders of magnitude due to beaming (Kulkarni
\etal 1999; M\'{e}sz\'{a}ros \& Rees 1999), although the relatively low
efficiency for converting internal shock energy into radiation might imply
an energy budget as high as $10^{54}$ ergs even after the beaming
correction (Kumar 1999).  Unless otherwise noted, we adopt the value
$E=10^{52}$ ergs, which is intermediate for the range inferred by Freedman
\& Waxman (1999). In \S~\ref{pardep} we will discuss the effect of beaming
and different choices of the explosion energy on our results.

\subsection{Non-Relativistic Regime}

As the blast wave decelerates, the fireball eventually makes a transition
from relativistic to sub-relativistic ($sr$) expansion (Huang, Dai \& Lu
1998; Woods \& Loeb 1999; Frail, Waxman \& Kulkarni 1999; Wei \& Lu
1999). In this regime, the evolution of the GRB remnants is well described
by the Sedov-Taylor solution (Taylor 1950; Sedov 1959). The time when this
transition takes place in the observer frame, $t_{sr}$, can be
approximately derived by setting the shock velocity in the Sedov-Taylor
solution to be equal to the speed of light, $c$.  This gives:
\begin{equation}
t_{sr} = \left(1+z\right)\left [\frac{E}{m_p n} \left ( \frac{0.47}{c}
\right )^5 \right ]^{1/3} = 1.8 \times 10^{2} (1+z) E_{52}^{1/3} n_1^{-1/3}
\;\;\; {\rm days}.
\label{tsr}
\end{equation}
In the sub-relativistic regime, the shock radius and velocity evolve as,
\begin{equation}
r(t)=r(t_{sr}) (t/t_{sr})^{2/5},
\label{rsr}
\end{equation}
\begin{equation}
v(t)=c (t/t_{sr})^{-3/5},
\label{vsr}
\end{equation}
where all times are in the observer frame and $r(t_{sr})$ is the shock
radius at time $t_{sr}$.  For a strong non-relativistic shock, the
post-shock particle density, energy density, and magnetic field are,
\begin{equation}
n^\prime=4n,
\end{equation}
\begin{equation}
u^\prime=\frac{9}{8} n m_p v^2,
\end{equation}
\begin{equation}
B^\prime=(8 \pi \epsilon_B u^\prime)^{1/2}.
\end{equation}
The observed flux requires a substitution of the values of $\nu_c$, $\nu_m$
and $F_{\nu _m}$ in the sub-relativistic regime into equations~(\ref{fnuf})
and~(\ref{fnus}). The synchrotron formulae can be used since the emitting
electrons are still ultra-relativistic for the relevant remnant ages in
this regime. The characteristic synchrotron frequency is given by,
\begin{equation}
\nu(\gamma)=\gamma^2 \frac{eB^\prime}{2 \pi m_e c}.
\label{nu}
\end{equation}
Recalling that $\gamma_m \propto B^{\prime 2} \propto u^\prime/n^\prime$,
we find $\gamma_m \propto v^2 \propto t^{-6/5}$, and $\gamma_c \propto
(B^{\prime 2} t)^{-1} \propto t^{1/5}$. Equation~(\ref{nu}) then provides
an expression for $\nu_c$ and $\nu_m$ in the sub-relativistic regime. In
particular $\nu_c \propto t^{-1/5}$ and $\nu_m \propto t^{-3}$.  During the
relativistic expansion the observed peak flux is constant in time; however,
in the sub-relativistic regime it varies as $F_{\nu_m} \propto r^3 B^\prime
\propto t^{3/5}$.

In the UV regime one needs to take account of absorption by the
intergalactic medium (IGM).  At redshifts greater than the reionization
redshift, $z_{reion}$, the neutral IGM is optically thick to photon
energies above the Ly$\alpha$ resonance.  Recent models for the
reionization of the IGM predict $z_{reion}$ in the range of 7-12 (Haiman \&
Loeb 1997; Miralda-Escud\'e, Haehnelt \& Rees 1998; Valageas \& Silk 1999;
Chiu \& Ostriker 1999; Gnedin 1999; Ciardi \etal 2000). Here we adopt
$z_{reion}=8$; the precise choice of which has only a minor effect on our
results since the Ly$\alpha$ forest yields a high opacity at the redshifts
of interest even if most of the IGM is ionized. For $z\la 5$, we use the
updated Ly$\alpha$ forest opacity derived by Haardt \& Madau (1996) and
Madau, Haardt \& Rees (1999), which is based on the observed absorber
distribution in the spectra of high-redshift quasars.  The continuum
depression blue-ward of the Ly$\alpha$ resonance is already close to unity
at $z\sim 5$ (see Figure~13 in Stern \& Spinrad 1999), and so the treatment
of the Ly$\alpha$ absorption at $5\la z\la 8$ has little effect on our
results.

Next, we derive an expression for the ambient gas density as a function of
redshift. For simplicity, we assume that the ambient gas has the mean
density of the interstellar medium of the host galaxy, as inferred for
GRB970508 and GRB971214 (Waxman 1997; Wijers \& Galama 1999).

\subsection{Evolution of the Gas Density in Galactic Disks}
\label{hge}

The phenomenological modeling of many afterglow lightcurves implies values
of the ambient medium density $\sim 1~{\rm cm}^{-3}$, which are of order
the mean density of the interstellar medium of disk galaxies (Waxman 1997;
Wijers \& Galama 1999; M\'esz\'aros 1999).  In popular hierarchical models
of galaxy formation (Kauffmann, White \& Guiderdoni 1993; Mo, Mao \& White
1998), the mass and size of galactic disks evolves with redshift (Barkana
\& Loeb 2000).  Hence, in modeling the statistical properties of GRBs
within disk galaxies at different redshifts, we need to follow the
evolution of the average density of the interstellar gas within these
galaxies as a function of cosmic time.

We model a disk galaxy as a gaseous disk embedded within a dark matter
halo.  The disk is assumed to be self-gravitating and radially exponential
with a scale radius, $r_d$.  Its scale height at any radius is dictated by
the balance between its self-gravity and the gas pressure. For simplicity,
we assume the disk to be isothermal. Its mass density profile is then
(Spitzer 1942),
\begin{equation}
\rho(\zeta,r)=\rho_0 {\rm e}^{-r/r_d} {\rm sech}^2 (\xi/\sqrt{2}),
\label{diskd}
\end{equation}
where $\zeta$ and $r$ are the vertical and radial coordinates of the disk,
respectively; $\rho_0=$const, $\xi=\zeta (4 \pi G \rho_0 {\rm e} ^{-r/r_d}/
c_s^2)^{1/2}$, and $c_s$ is the effective sound speed (including the
possible contribution from turbulence).  We adopt a sound speed of $c_s
\approx 10$~km~s$^{-1}$, corresponding to a temperature of $\sim 10^4~{\rm
K}$ below which atomic cooling is highly inefficient. At redshifts $z\la
15$ the formation of galaxies with virial temperatures below $10^4~{\rm K}$
is expected to be suppressed due to various feedback effects (Barkana \&
Loeb 1999; Ciardi \etal 2000).

The scale-radius of the disk is dictated by the angular momentum per
unit mass of the gas and can be expressed as
(Mo, Mao \& White 1998),
\begin{equation}
r_d=\frac{1}{\sqrt{2}} \frac{j_d}{m_d} \lambda r_{vir}.
\end{equation}
Here $j_d=J_d/J$ and $m_d=M_d/M$, where $J$ and $M$ are the halo angular
momentum and mass and the subscript $d$ refers to the disk; $\lambda$ and
$r_{vir}$ are the halo spin parameter and virial radius respectively. We
assume that the disk mass fraction is the cosmic baryonic fraction,
$M_d/M=(\Omega_b/\Omega_0)$, since gas cooling is efficient at the high
redshifts of interest.
The characteristic size distribution of
local galactic disks is obtained by adopting $j_d/m_d=1$ (Fall \&
Efstathiou 1980).  Based on numerical simulations, we parametrize the
distribution of the spin parameter $\lambda$ by 
a log-normal form (Mo, Mao \& White 1998, and
references therein),
\begin{equation}
p(\lambda) d\lambda= \frac{1}{\sigma_\lambda \sqrt{2 \pi}} {\rm exp}
\left [- \frac{{\rm ln}^2 (\lambda / \overline{\lambda})}{2 \sigma_\lambda^2}
\right ] \frac{d \lambda}{\lambda},
\end{equation}
with $\overline{\lambda}=0.05$ and $\sigma_\lambda=0.5$.  The halo virial
radius is given by (Barkana \& Loeb 1999),
\begin{equation}
r_{vir}=7.56 \left [ h^{-2} \frac{M}{10^8 M_\odot} \frac{\Omega (z)}
{\Omega_m} \frac{200}{\delta_c} \right ]^{1/3}
(1+z)^{-1} \; {\rm kpc},
\label{rvir}
\end{equation}
with $\delta_c=18 \pi^2$ and,
\begin{equation}
\Omega (z)=\frac{\Omega_m (1+z)^3}{\Omega_m (1+z)^3+\Omega_\Lambda+
(1-\Omega_m-\Omega_\Lambda)(1+z)^2}.
\end{equation}
The total mass of the disk, $M_d$, can be derived by integrating equation
(\ref{diskd}). This yields the relation,
\begin{equation}
\rho_0=\frac{G M_d^2}{128 \pi c_s^2 r_d^4}.
\end{equation}

The star formation rate depends on the local surface density of the gaseous
disk.  Schmidt's law implies that the star formation rate is proportional
to the local surface density of the disk to the power of 1-2 (Kennicutt
1998).  
Hence we first model the probability for a GRB occurrence within a
volume element $2 \pi r dr d\zeta$ as being proportional to the density
squared, $P_{GRB}= {\cal A} n^2(\zeta,r)$; while in \S 3.3 we consider the
sensitivity of our results to other power-laws scalings. The
proportionality constant ${\cal A}$ is chosen to normalize the integral of
$P_{GRB}$ to unity.  Consequently, we can write the average of the ambient
density probed by GRBs as
\begin{equation}
\langle \rho \rangle = \int_0^\infty \int_0^\infty \rho (\zeta,r) P_{GRB} 2
\pi r dr d\zeta = 0.29 \rho_0 .
\label{rhom}
\end{equation}
The number density of atoms in the medium is thus $\langle n \rangle =
\langle \rho \rangle/ \mu m_p$, where $\mu$ is the mean molecular
weight. Similarly, we can derive $\langle \sqrt{n} \rangle$ and $\langle
1/n \rangle$ and substitute them into equations~(\ref{nuc})-(\ref{fmax}).
Even if the central engine of all GRBs were standard, the dependence of the
ambient medium density on the properties (mass and size) of the host galaxy
would have introduced significant scatter and evolution to the flux
distribution of GRB afterglows.

\section{Results}\label{res}

\subsection{Flux Evolution}

\begin{figure}[t]
\plotone{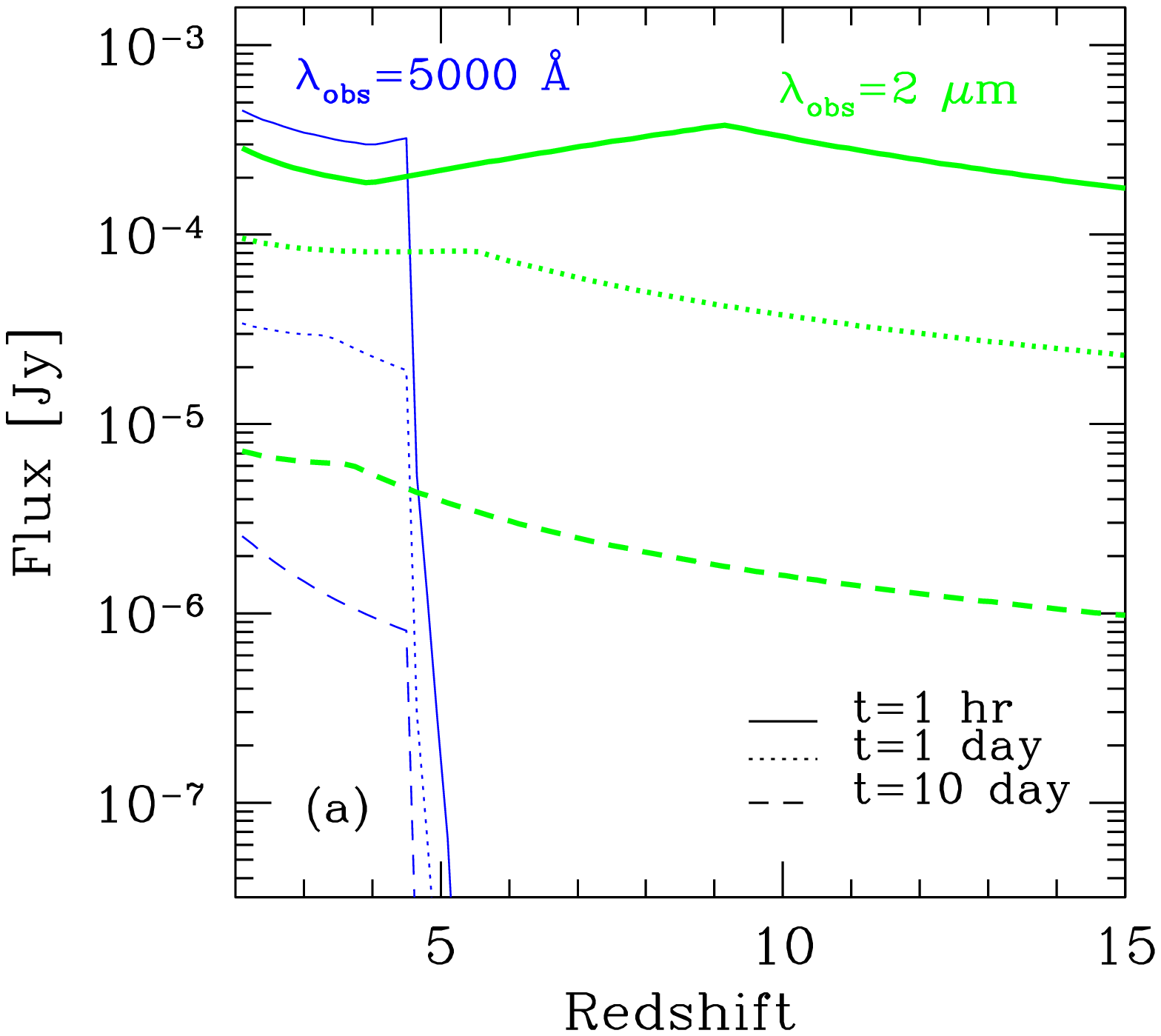}
\caption{\label{fig1}{(a) Observed flux for a GRB hosted by a ''typical''
halo (having the mean mass) as a function of redshift (see text for
details).  The curves are for a frequency $\nu=6 \times 10^{14}$ Hz
($\lambda_{obs}=5000$ \AA, thin lines) and $\nu=1.5 \times 10^{14}$ Hz
($\lambda_{obs}=2 \mu$m, thick lines). From the top to the bottom different
observed times after the GRB are shown: one hour (solid line), 1 day
(dotted) and 10 days (dashed).  }}
\end{figure}

Figure~\ref{fig1} shows the observed flux for a GRB hosted by the average
halo mass as a function of redshift.  This average is calculated based on
the Press \& Schechter (1974) formalism, which provides the mass function
of dark matter halos as a function of redshift. The two sets of curves
correspond to two observed wavelengths: in Figure~\ref{fig1}a the thick
lines correspond to $\lambda_{\rm obs}=2\mu$m and the thin lines to
$5000$\AA, while in Figure~\ref{fig1}b the thick lines correspond to
$\lambda_{\rm obs}=10$ cm and the thin lines to 1 mm. Within each set, the
lines correspond to an observed time of 1 hr (solid), 1 day (dotted) and 10
days (dashed).  While the optical and infrared emission peak at observed
times shorter than an hour, the radio emission has not reached its maximum
even after 50 days.

At observed frequencies below the Ly$\alpha$ resonance frequency,
$\nu_\alpha(z)=2.47\times 10^{15}/(1+z)$ Hz, the emitted flux is not
absorbed by the intervening intergalactic medium at a redshift $z$. 
Figure~\ref{fig1} shows that at these frequencies the afterglow flux is only 
weakly dependent on redshift. This can be crudely understood from the scaling 
laws implied by equations~(\ref{fnuf}) and~(\ref{fnus}), by considering the
simple Einstein--de Sitter cosmology ($\Omega_m=1$), for which $d_L \propto
(1+z) [1-(1+z)^{-1/2}]$. Recalling that $n \propto (1+z)^4$ for a fixed
host galaxy mass [see eqs.~(\ref{rvir})-~(\ref{rhom})] and
substituting these scaling laws into equations~(\ref{fnuf})
and~(\ref{fnus}), we find for the case of fast cooling,
\begin{equation}
F_\nu \propto \left\{
\begin{array}{ll}
(1+z)^{5/2} [1-(1+z)^{-1/2}]^{-2} & \nu_c>\nu,\\
(1+z)^{-5/4} [1-(1+z)^{-1/2}]^{-2} & \nu_m>\nu \geq \nu_c,\\
(1+z)^{(p-6)/4} [1-(1+z)^{-1/2}]^{-2} & \nu \geq \nu_m;\\
\end{array}\right.
\end{equation}
while for slow cooling,
\begin{equation}
F_\nu \propto \left\{
\begin{array}{ll}
(1+z)^{5/6} [1-(1+z)^{-1/2}]^{-2} & \nu_m>\nu,\\
(1+z)^{(p-3)/4} [1-(1+z)^{-1/2}]^{-2} & \nu_c>\nu \geq \nu_m,\\
(1+z)^{(p-6)/4} [1-(1+z)^{-1/2}]^{-2} & \nu \geq \nu_c.\\
\end{array}\right.
\end{equation}
Although these scaling laws ignore the evolution of the characteristic
galaxy mass with redshift, they demonstrate that the observed flux does not
decline rapidly with increasing redshift, but might rather increase with
redshift at low frequencies (e.g. in the millimeter or radio regimes).
This behavior results from three causes: (i) for a fixed observed time-lag
after the GRB, the cosmic time-dilation implies that the higher the source
redshift is, the earlier the emission time is, and the brighter the
intrinsic afterglow luminosity is; (ii) the spectral slope of GRB
afterglows results in a favorable K-correction for the redshifts and
wavelengths of interest; and (iii) the gas density in galaxies increases
with redshift (due to the corresponding increase in the mean density of the
universe).

\subsection{Number Count}
\label{nc}

In a snapshot mode of observations, the total number of GRBs with observed
flux greater than $F$ at an observed frequency $\nu$ is
\begin{equation}
N(>F,\nu) = \int_0^\infty \int_{M_{min}}^{\infty} r_{GRB}(z,M)
\frac{t_F(z,M,\nu)}{(1+z)} \frac{dn}{dM}(z) \frac{dV}{dz} dM dz.
\label{nf}
\end{equation}
where $r_{GRB}$ denotes the GRB rate in a galaxy halo of mass $M$ at a
redshift $z$, $t_F$ is the observed time over which a GRB event from this
halo yields a flux brighter than $F$ at a frequency $\nu$, $dn/dM$ is the
comoving number density of halos with masses between $M$ and $M+dM$ (based
on the Press-Schechter formalism), and $M_{min}$ is the minimum galaxy mass
in which stars form at a redshift $z$.  The minimum halo mass inside which
star formation occurs, is related to a minimum virial temperature of $\sim
10^4~{\rm K}$ below which atomic cooling of the gas is suppressed and
fragmentation into stars is inhibited (Haiman, Rees \& Loeb 1997; Ciardi
\etal 2000). This minimum virial temperature implies $M_{\rm min}=4.4
\times 10^9 M_\odot (1+z)^{-1.5} h^{-1}$ (Padmanabhan 1993), and is also
consistent with our choice for the sound speed of the disk (see
\S~\ref{hge}).  The comoving volume element per unit redshift, $dV/dz$, is
given by,
\begin{equation}
\frac{dV}{dz}=\frac{4 \pi c d_L^2}{(1+z)} \left\vert \frac{dt}{dz}
\right\vert,
\end{equation}
where $d_L$ and $dt/dz$ are given in equations~(\ref{dl}) and~(\ref{dtdz}),
respectively. Next we consider the value of the GRB rate, $r_{GRB}$.

If GBRs are related to the final stages in the evolution of massive stars,
then one may assume that the GRB rate, $r_{GRB}$, is proportional to the
star formation rate (SFR). This assumption is justified as long as the time
delay between the formation of a massive star and a GRB event is short
compared to the Hubble time at the redshift of interest.  We adopt this as
our working hypothesis.  Wijers \etal (1998) have estimated the constant of
proportionality between $r_{GRB}$ and the cosmic SFR for an Eistein-de
Sitter cosmology, but their results do not change significantly for other
values of the cosmological parameters (Bagla 1999, private
communication). Using the observed redshift history of the SFR per comoving
volume in the universe (Lilly \etal 1996; Madau \etal 1996) and the
observed distribution of $\gamma$-ray flux of GRBs, Wijers et al. (1998)
have calibrated the GRB rate per comoving volume.  Assuming that GRBs are
standard candles, they have found a GRB rate at $z=0$ of $R_{GRB}(0)=0.14
\pm 0.02$ Gpc$^{-3}$ yr$^{-1}$.  This estimate ignores recent corrections
to the cosmic SFR. The same calculation, performed with a SFR per comoving
volume that flattens at high redshift, as suggested by more recent
corrections for dust extinction, still leads to similar results (Bagla
1999, private communication). Assuming that the proportionality constant
does not vary with redshift, we write,
\begin{equation}
r_{GRB}(z,M) = \frac{R_{GRB}(0)}{\dot{\rho}_\star(0)} \dot{M}_\star(z,M).
\end{equation}
Here $\dot{\rho}_\star(0)$ is the star formation rates per comoving volume
in units of M$_\odot$ yr$^{-1}$ Mpc$^{-3}$, and $\dot{M}_\star
(z,M)=\alpha(z) \Omega_b M$ is the SFR within a particular halo in units of
M$_\odot$ yr$^{-1}$.  We calibrate the star formation efficiency as a
function of redshift, $\alpha(z)$, so as to match the observed cosmic SFR,
$\dot{\rho}_\star$, at $z \la 4$,
\begin{equation}
\dot{\rho}_\star(z)=\int_{M_{min}}^{\infty} \dot{M}_\star(z) \frac{dn}
{dM}(z) dM.
\label{sfr}
\end{equation}
An analytical fit for the observed $\dot{\rho}_\star(z)$ (assuming a
Salpeter IMF and an extinction correction of A$_{1500}=1.2$ mag) 
was derived by Madau \& Pozzetti (1999),
\begin{equation}
\dot{\rho}_\star(z)=\frac{0.23 {\rm e}^{3.4z}}{{\rm e}^{3.8z}+44.7}
\;\;\; {\rm M_\odot yr^{-1} Mpc^{-3}}.
\end{equation}
At $z>4$ the SFR within a particular halo is calculated using the common
prescription $\alpha(z) \propto t_{dyn}^{-1}(z)$, where $t_{dyn}(z)$ is the
mean dynamical time of the halo.  The constant of proportionality is chosen
so as to match the observed cosmic SFR at $z=4$ (see Ciardi et al. 2000,
for more details).  

Figure~\ref{fig2} presents the total number counts based on
equation~(\ref{nf}). The set of curves corresponds to different observed
wavelengths. From the right to the left, $\lambda_{obs}$ is equal to 10 cm,
1 mm, 2 $\mu$m and 5000 \AA. At low fluxes the number counts approach an
asymptotic value.  Out of the entire population of faint GRB afterglows, we
predict that there are $\sim 15$ GRBs from redshifts $z\ga 5$ across the
sky which are brighter than $\sim 100$ nJy at an observed wavelength of
$\sim 2\mu$m.  Figure~\ref{fig3} illustrates the contribution to the total
number counts from different redshift bins, centered at $z$=2 [Fig.~(3a)],
6 [Fig.~(3b)] and 10 [Fig.~(3c)].  Infrared or radio afterglows are
detectable out to redshifts as high as 10, while optical afterglows are
strongly absorbed by the intervening IGM at $z\ga 5$.  Figure~\ref{fig4}
shows the differential number count distribution per logarithmic flux
interval and redshift interval of all sources at a given redshift.  At high
frequencies, for which the observed flux decreases with increasing
redshift, low-redshift events dominate the counts at most fluxes.  However,
at low frequencies (\ie millimetric wavelength) the high-redshift events
dominate the counts at sufficiently low fluxes.

\subsection{Dependence of Results on Model Assumptions}
\label{pardep}

Next we consider the dependence of our results on various model parameters.
Figure~\ref{fig5}a shows the observed flux at $\lambda_{obs}=2\mu$m and
$t=1$ day for different choices of the mean ambient density. The standard
case with $P_{GRB}\propto n^2$ (solid line) is compared to two other cases;
one where $P_{GRB}\propto n$ (dot-dashed line) and the other where the
ambient density is kept constant and equal to 1 cm$^{-3}$ (dotted line).
The figure indicates that for observations in the infrared, all three cases
provide identical results at $z \ga 5$, when one enters a regime where the
observed flux becomes independent of the ambient density [either the fast
cooling regime with $\nu>\nu_m$ or the slow cooling regime with
$\nu>\nu_c$, given by Eqs.~(\ref{fnuf}) and~(\ref{fnus})].  Outside this
regime the observed flux is roughly proportional to $n^{5/6}$ and
$\sqrt{n}$ in the fast and slow cooling cases, respectively. In the slow
cooling regime, for example, the condition $\nu \ge \nu_c$ leads to a
minimum redshift above which the flux is independent of ambient density
$(1+z) \geq 0.32 E_{52}^{-1} n_1^{-2} t_d^{-1} \left({\lambda_{\rm obs}/ 2
\mu{\rm m}}\right)^{2}$. Hence, at long wavelengths or very low densities,
the density independent regime is not reached.  Figure~\ref{fig5}a shows
that the difference between the $P_{GRB} \propto n^2$ and $P_{GRB} \propto
n$ cases, is small since they follow a similar evolution of the gas density
with redshift.  Figure~\ref{fig5}b shows the number count of GRBs per
logarithmic redshift interval at $z=6$, both for the standard case with
$P_{GRB} \propto n^2$ (solid line) and the case of a constant ambient
density equal to 1 cm$^{-3}$ (dotted line). The difference between these
cases is small.

Figure~\ref{fig6}a shows the dependence on the energy output of the
afterglow number count from a redshift bin centered at $z=6$ for an
observed wavelength of 2$\mu$m.  We consider two cases that bracket the
standard case ($E=10^{52}$ ergs); namely $E=10^{53}$ erg (dotted) and $E=5
\times 10^{50}$ erg (dashed). The increase in the observed flux as the
energy output increases [see eqs.~(\ref{fnuf})--(\ref{fmax})], is reflected
in the afterglow number count.

So far we have assumed that the GRB explosions are spherically-symmetric.
If, however, the energy release is beamed, then the afterglow lightcurve is
expected to evolve differently than we assumed.  At early times, as long as
the expansion Lorentz factor, $\gamma$, is still larger than the inverse of
the jet opening angle, $\theta^{-1}$, the expansion behaves as if it were
spherically symmetric with the same energy output per solid angle.  Once
the jet decelerates to a Lorentz factor, $\gamma\la \theta^{-1}$, the
lightcurve declines more rapidly, since the jet starts to expand sideways
and to reduce the mean energy output per solid angle (Rhoads 1999a,b;
Panaitescu \& Meszaros 1999). Finally, the isotropization of the energy
ends when the expansion becomes sub-relativistic, at which time the
lightcurve recovers the spherically-symmetric Taylor-Sedov evolution for
the actual total energy output.  (Synchrotron radiation losses are usually
negligible in the late afterglow phase.)  For an observing point which is
aligned with the GRB jet axis, the afterglow lightcurve may start with an
isotropic-equivalent energy of $E$ and end in the non-relativistic regime
with the actual energy output of $E_{sr}=\eta E$ where
$\eta=(\Delta\Omega/4\pi)=2(\pi \theta^2/4\pi)\leq 1$ is the fraction of
sky around the GRB source which is illuminated by two opposing jets of
angular radius, $\theta$.  Hence, the afterglow makes a transition between
the lightcurve corresponding to a highly-relativistic explosion with a
total energy $E$ to the lightcurve of a non-relativistic explosion with an
energy $\eta E$.  To examine the effect of beaming on our results, we model
this transition using a linear interpolation between the two lightcurves.
We start with an isotropic equivalent of $E=10^{53}~$ergs, and adopt
$\theta=0.1$ so that the actual energy output is $E_{sr}=5 \times 10^{50}$
erg (since $\gamma\ga 10^2\gg \theta^{-1}$, initially). As expected, the
afterglow fluxes and source number counts make a transition between their
values for the two boundary energies.  Figures~\ref{fig6}a-c show the
transition between the two regimes with a dot-dashed line.  We find that,
for $\lambda_{\rm obs}=2\mu$m, the number counts are dominated by sources
in the relativistic phase. However, at longer wavelengths,
$\lambda_{obs}=1$ mm or 10 cm [Fig.~\ref{fig6}b or Fig.~\ref{fig6}c,
respectively], the bulk of the emission takes place at late times when the
expansion is in its mildly or sub-relativistic phase.  Obviously, the
effect of beaming can be recovered from the shape of the afterglow
lightcurve in individual GRB events.  Figure~\ref{fig6} shows that the
existence of beaming might also be inferred from the number count
statistics.

\section{Discussion}\label{disc}

We have calculated the expected fluxes and number counts of high-redshift
GRB afterglows, assuming that their rate of occurrence is proportional to
the star formation rate. We have computed the observed properties of GRB
lightcurves at different wavelengths ranging from the optical to the radio,
treating both the relativistic and sub-relativistic phases of the expansion
of their remnants.

Our main result is that at a fixed observed time after the GRB event, the
characteristic afterglow flux is not decreasing rapidly with increasing
redshift (Figs. 1a, 1b), in contrast with other high-redshift sources,
such as galaxies or quasars.  Hence, the broad-band spectrum of GRB
afterglows is ideally suited for probing the ionization state and metal
content of the intergalactic medium at high redshifts, particularly during
the epoch of reionization (Loeb 1999; Lamb \& Reichart 1999).

The only difficulty in using GRBs as probes of the high-redshift universe
is that they are rare, and hence their detection requires surveys which
cover a wide area of the sky (see the vertical axis in
Figures~\ref{fig2}--\ref{fig4}).  The simplest strategy for identifying
high-redshift afterglows is through all-sky surveys in the $\gamma$-ray or
X-ray regimes.  In particular, detection of high redshift sources will
become feasible with the high trigger rate provided by the forthcoming {\it
Swift} satellite, to be launched in 2003.  {\it Swift} is expected to
observe and localize $\sim$300 GRBs per year, and to repoint within 20-70
seconds its on-board x-ray and UV-optical instrumentation for continued
afterglow studies. The high-resolution GRB coordinates obtained by {\it
Swift}, will be transmitted to Earth within $\sim$50 seconds. Deep
follow-up observations will then become feasible from the ground or using
the highly-sensitive infrared instruments on board the {\it Next Generation
Space Telescope} (NGST), scheduled for launch in 2008.  {\it Swift} is
sufficiently sensitive to trigger on the $\gamma$-ray emission from GRBs at
redshifts $z\ga 10$ (Lamb \& Reichart 1999).  We note that even if some
GRBs occur outside of galaxies, the intergalactic medium would be
sufficiently dense to produce an afterglow at these high-redshifts since
its mean density is $\ga 10^{-4} [(1+z)/10]^3~{\rm cm^{-3}}$.

The effect of beaming is expected to be minor for sources with millimeter
and radio fluxes $\la 10^{-5}$ Jy, since the number counts at these fluxes
are dominated by sources which are in the sub-relativistic phase of their
expansion, and for which the energy output in the explosion is already
isotropized (see Fig.~\ref{fig6}b,~\ref{fig6}c).

For a characteristic energy output of GRBs of $\sim 10^{52}$ ergs, our
model implies that at any time there should be $\sim 15$ GRBs from
redshifts $z \simgt 5$ across the sky, which are brighter than $100$
nJy at an observed wavelength of 2 $\mu$m.  The {\it Next Generation Space
Telescope} (NGST) will be able to measure the spectra of these sources.
Prior to reionization, the spectrum of GRB afterglows might reveal the long
sought-after Gunn-Peterson trough (Gunn \& Peterson 1965), due to
absorption by the neutral intergalactic medium.

\acknowledgments{We thank F. Haardt for providing us the IGM opacity
curves, and J. Bagla and R. Croft for useful discussions.  BC thanks the
CfA pre-doctoral fellowship program for support during the course of this
work. This work was also supported in part by NASA grants NAG 5-7039 and
NAG 5-7768, NSF grant 9900077, and by a grant from the Israel-US BSF.}

\newpage

\setcounter{figure}{0}

\begin{figure}[t]
\plotone{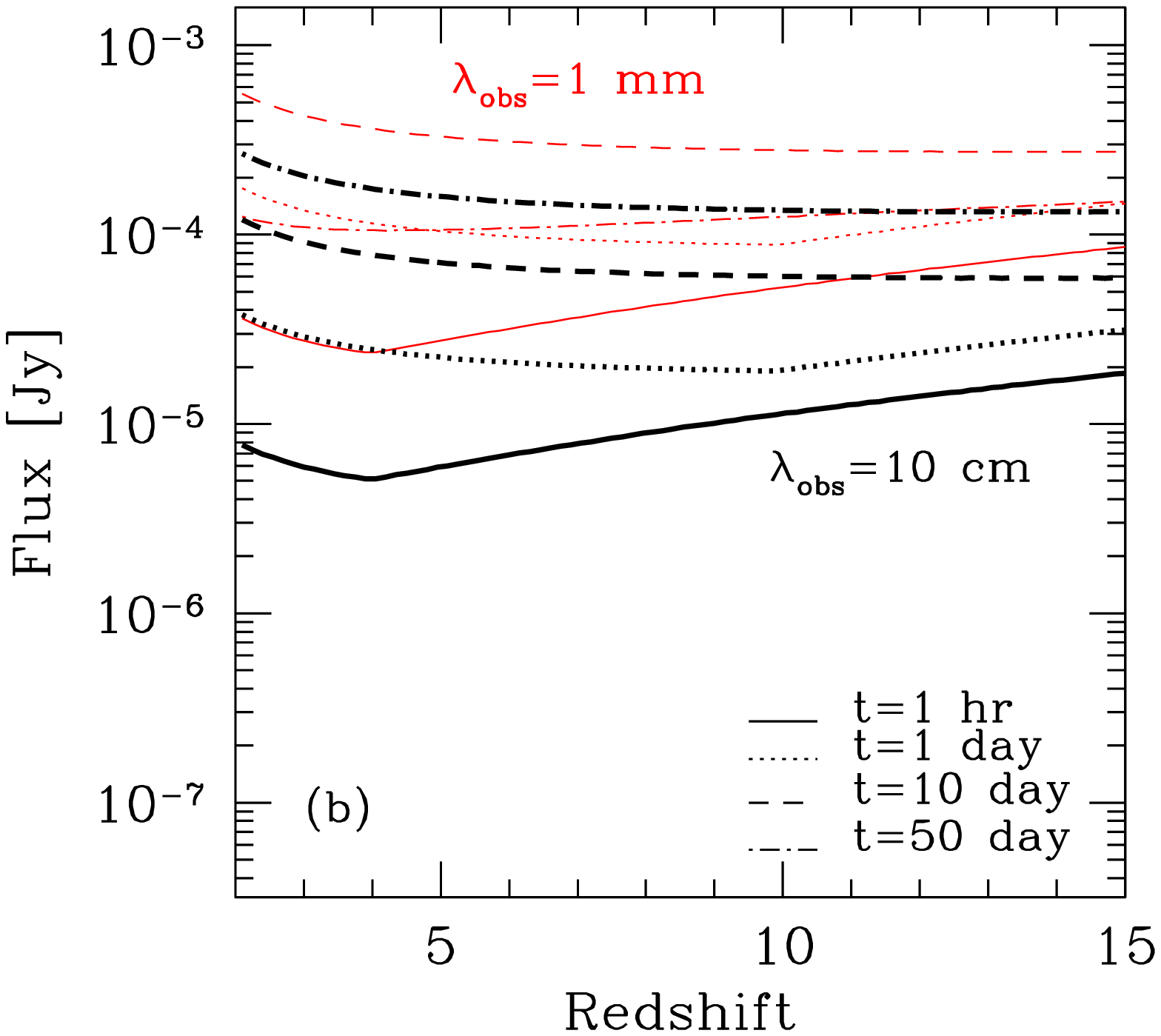}
\caption{{(b) As in Fig. (1a), but for $\nu=3 \times 10^{11}$ Hz
($\lambda_{obs}=1$ mm, thin lines) and $\nu=3 \times 10^9$ Hz
($\lambda_{obs}=10$ cm, thick lines). The dot-dashed line corresponds to an
observed time of 50 days.}}
\end{figure}

\begin{figure}[t]
\plotone{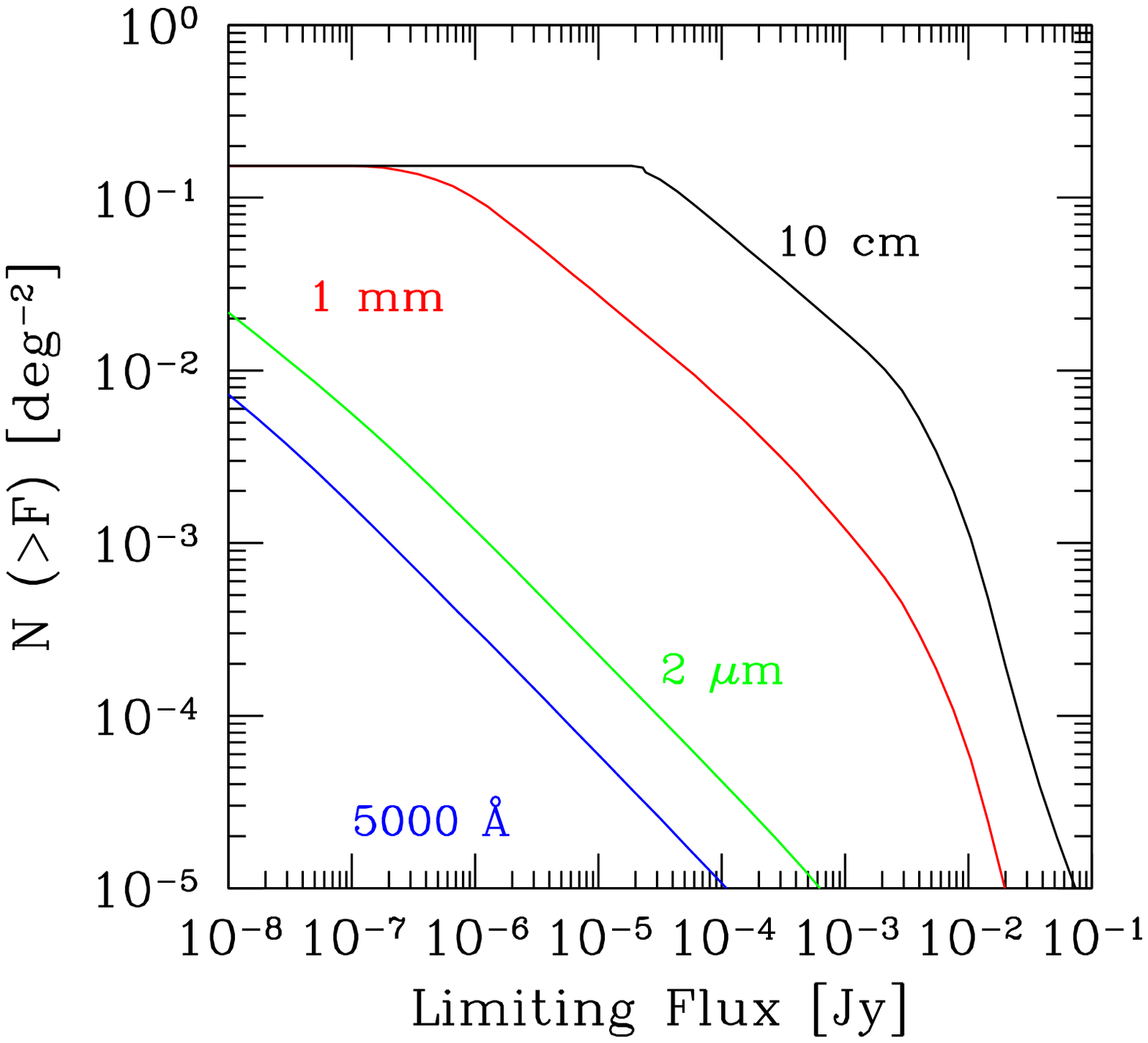}
\caption{\label{fig2}{Total number of GRBs with observed flux greater
than a limiting flux, $F$, at different observed wavelength
$\lambda_{obs}$.  From the right to the left $\lambda_{obs}$ is equal to 10
cm, 1 mm, 2 $\mu$m and 5000 \AA.}}
\end{figure}

\begin{figure}[t]
\plotone{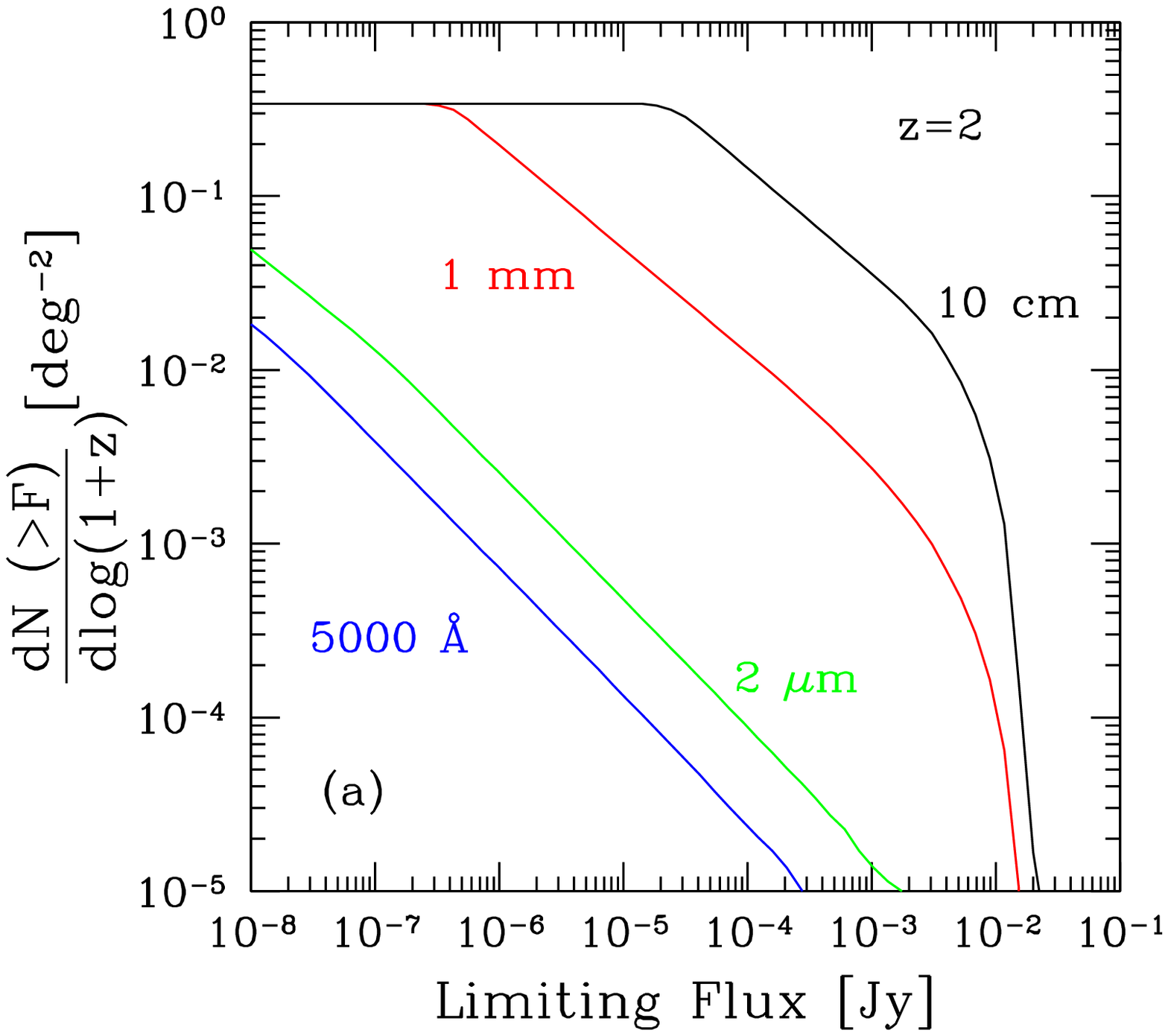}
\caption{\label{fig3}{(a) Contribution to the total number count
(Fig.~\ref{fig2}) of GRBs from a redshift bin centered at z=2. The observed
wavelengths are, from the right to the left: $\lambda_{obs}= 10$ cm, 1 mm,
2 $\mu$m and 5000 \AA.}}
\end{figure}
\setcounter{figure}{2}
\begin{figure}[t]
\plotone{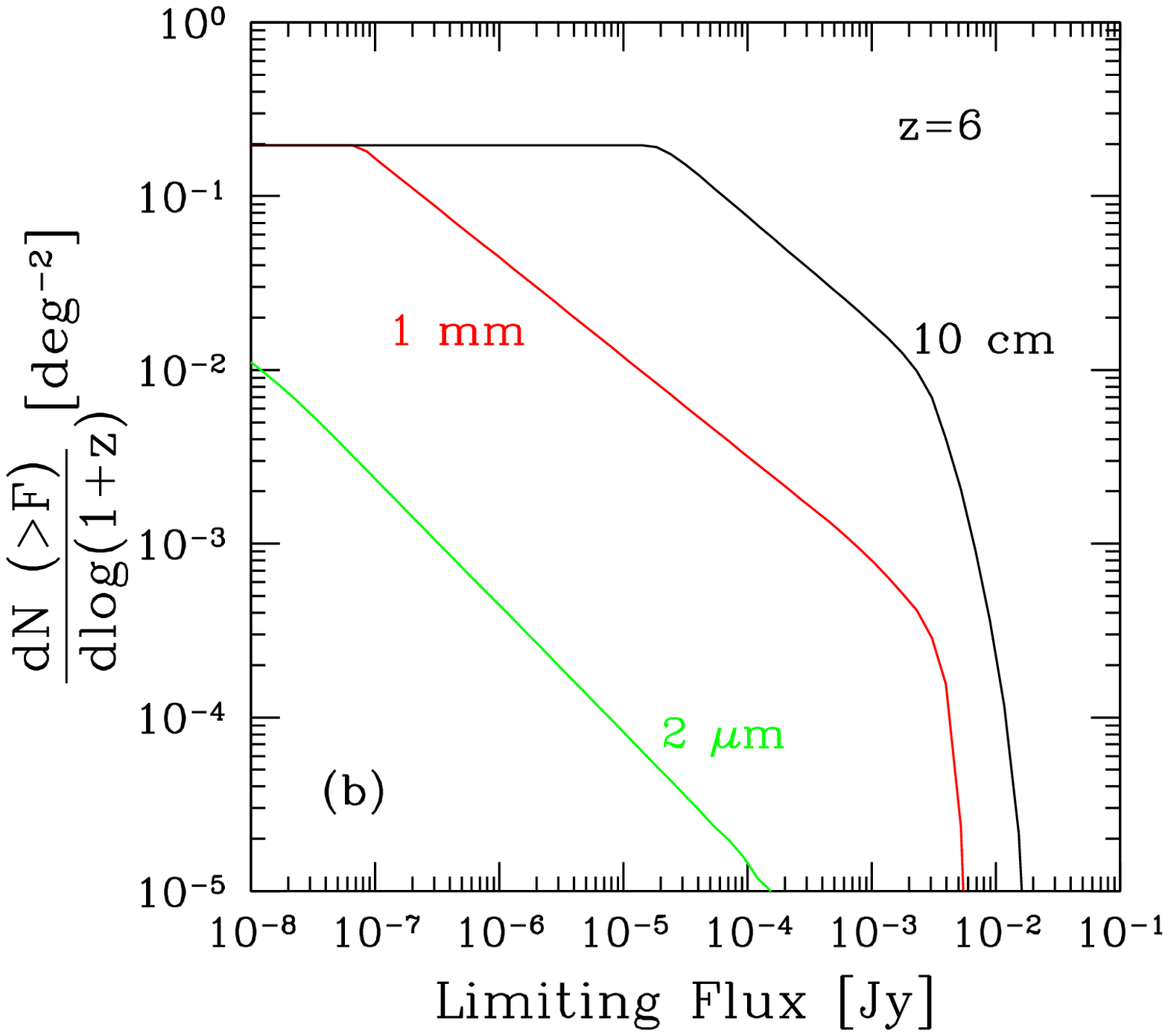}
\caption{{(b) As in Fig.~(3a) with a redshift bin centered at z=6.}}
\end{figure}
\setcounter{figure}{2}
\begin{figure}[t]
\plotone{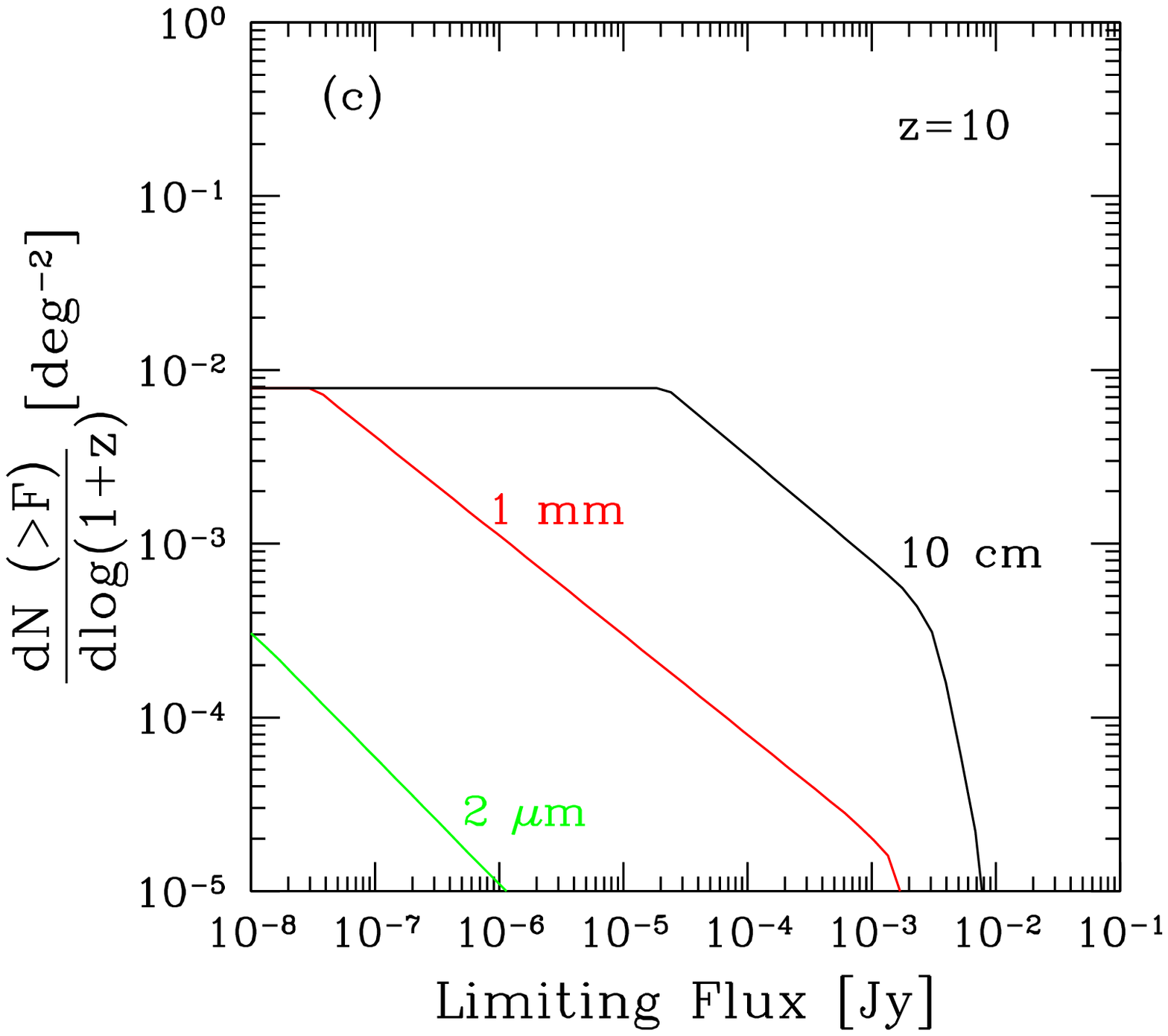}
\caption{{(c) As in Fig.~(3a) with a redshift bin centered at z=10.}}
\end{figure}

\begin{figure}[t]
\plotone{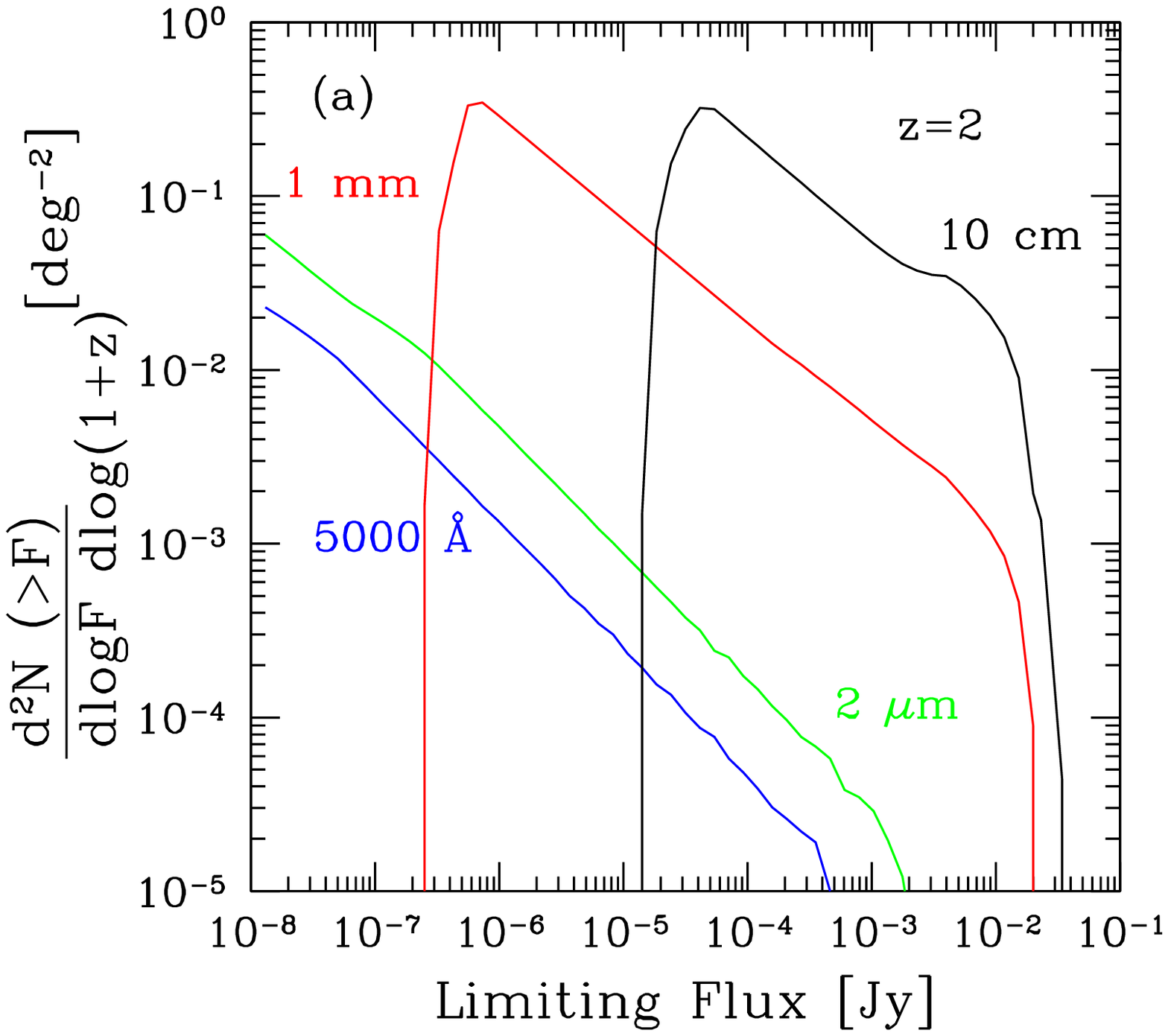}
\caption{\label{fig4}{(a) Number count distribution per logarithmic
flux and redshift interval. The observed wavelengths are, from the right to
the left: $\lambda_{obs}=10$ cm, 1 mm, 2 $\mu$m and 5000 \AA.  The redshift
bin is centered at z=2.}}
\end{figure}
\setcounter{figure}{3} 
\begin{figure}[t]
\plotone{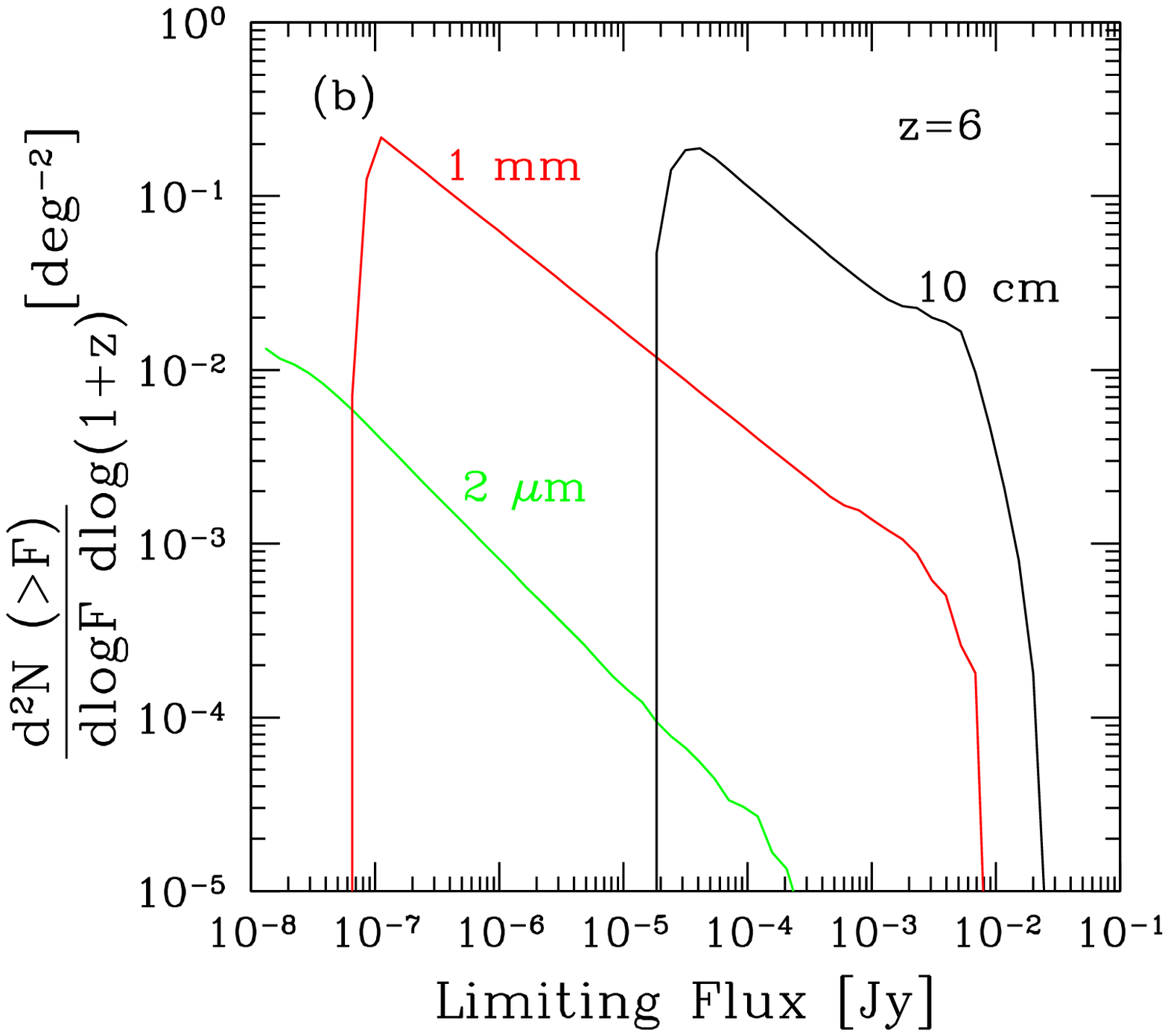}
\caption{{(b) As in Fig.~(4a) with a redshift bin centered at z=6.}}
\end{figure}
\setcounter{figure}{3} 
\begin{figure}[t]
\plotone{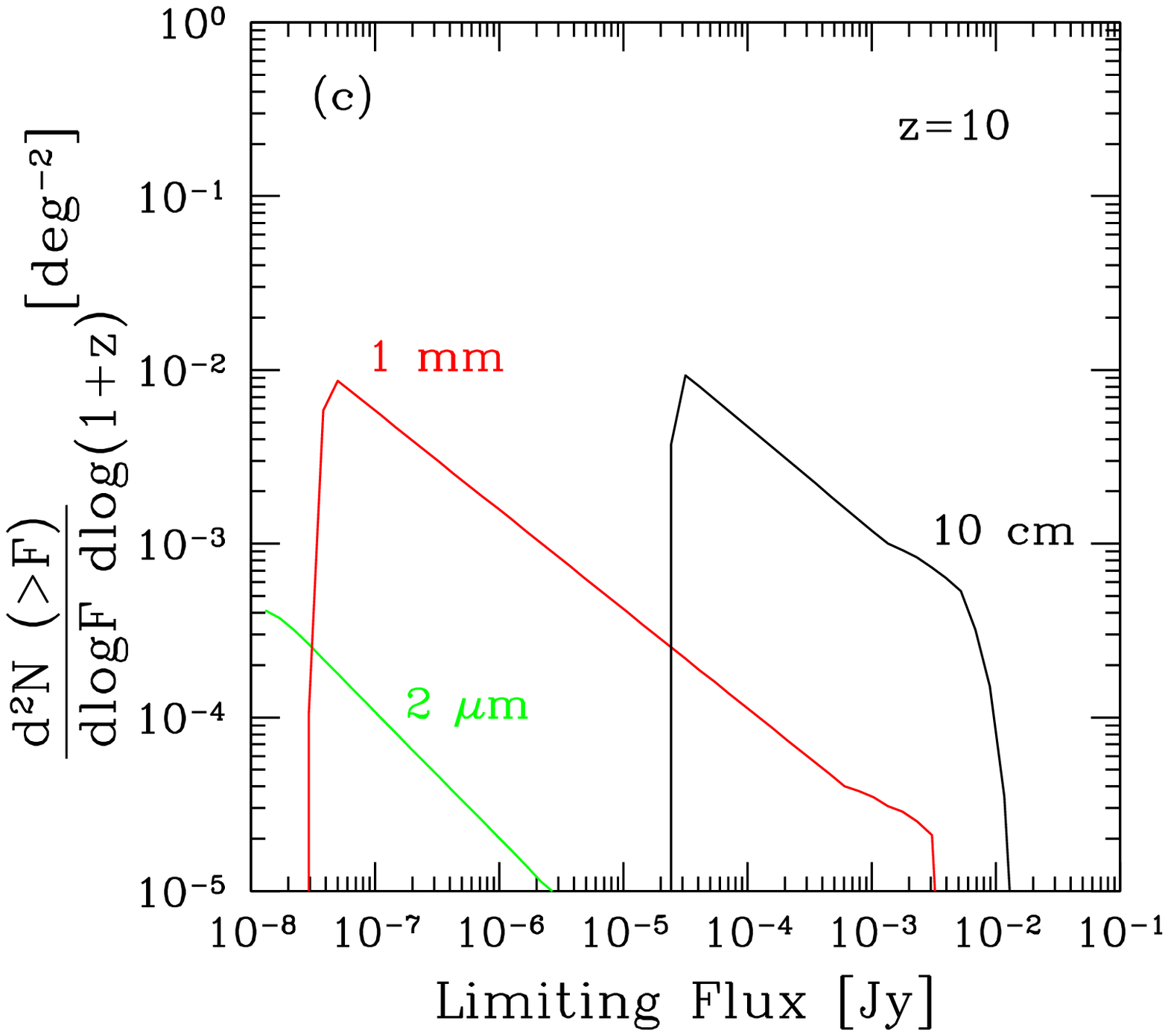}
\caption{{(c) As in Fig.~(4a) with a redshift bin centered at z=10.}}
\end{figure}

\begin{figure}[t]
\plotone{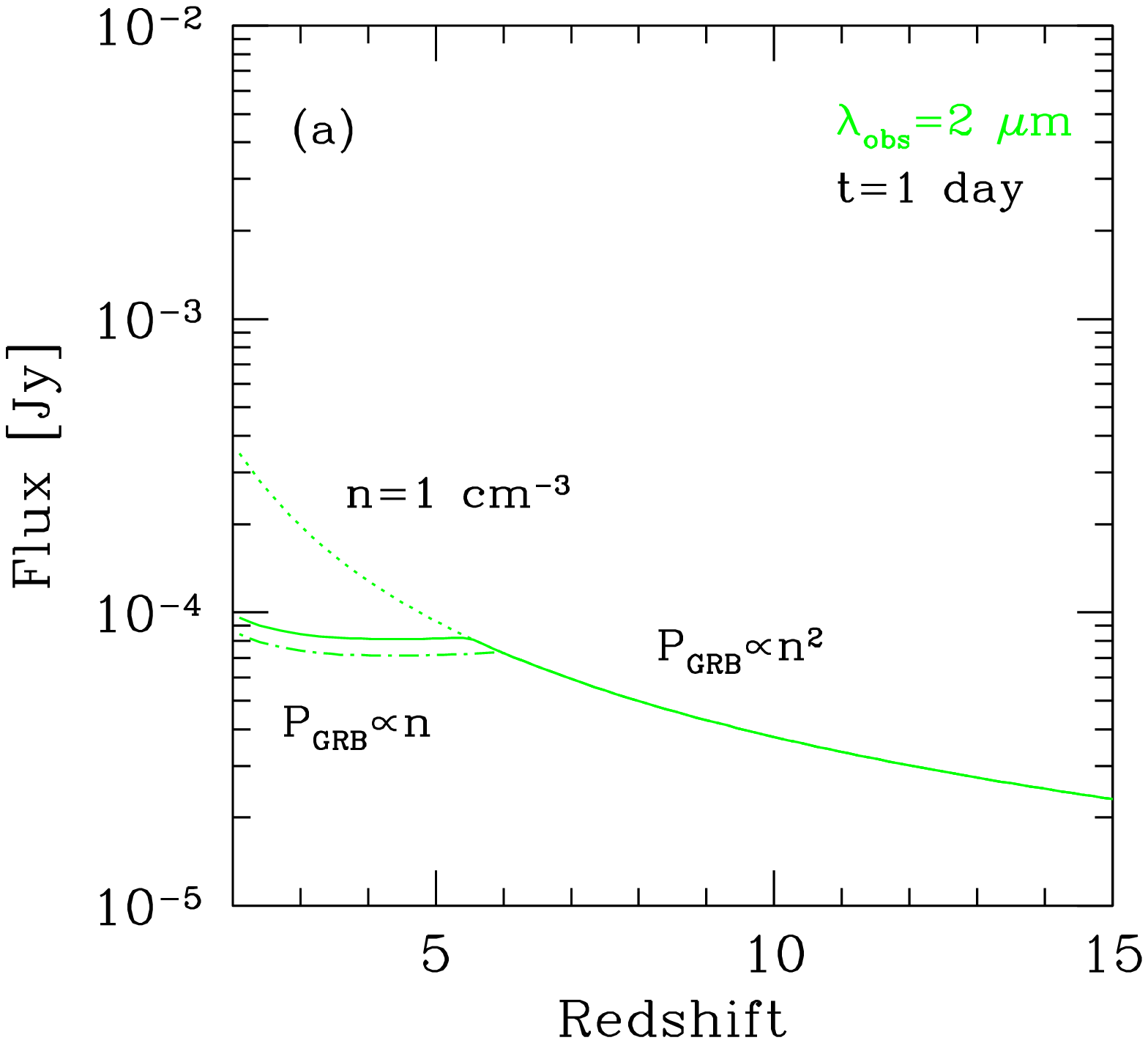}
\caption{\label{fig5}{(a) Observed flux for a GRB hosted by a ``typical''
halo (having the mean mass) as a function of redshift (see text for
details). The flux is shown for the reference observed wavelength 2 $\mu$m
and an observation time of 1 day after the GRB. The different curves
correspond to different ambient densities, namely: a constant density of 1
cm$^{-3}$ (dotted line), and density profiles of galactic disks with
$P_{GRB} \propto n^2$ (solid line; standard reference case) and with
$P_{GRB} \propto n$ (dot-dashed line).}}
\end{figure}
\setcounter{figure}{4}
\begin{figure}[t]
\plotone{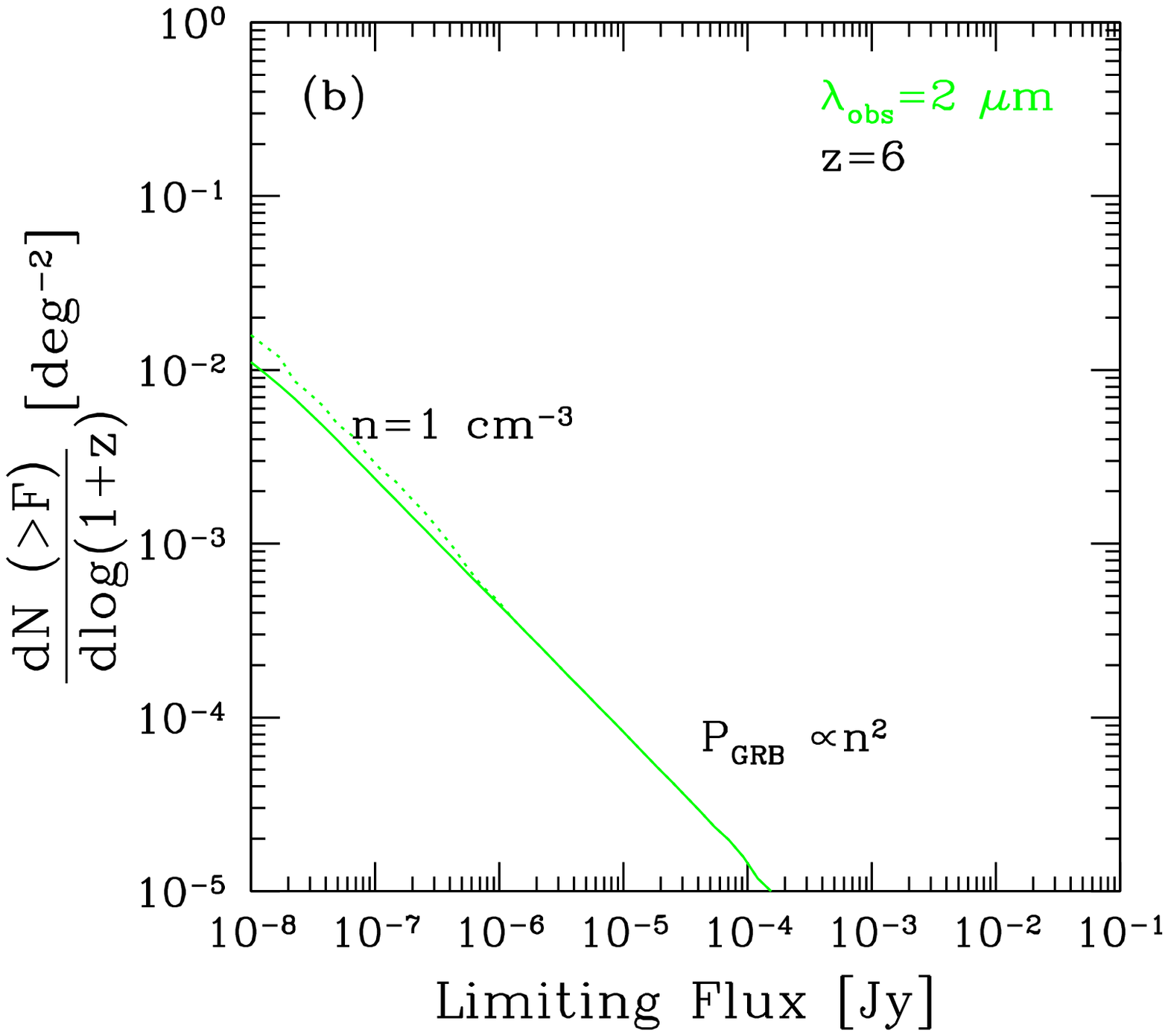}
\caption{{(b) Contribution to the total number count of GRBs from a
redshift bin centered at $z=6$ for an observed wavelength of 2
$\mu$m. Notation is the same as in Fig.~(5a).}}
\end{figure}

\begin{figure}[t]
\plotone{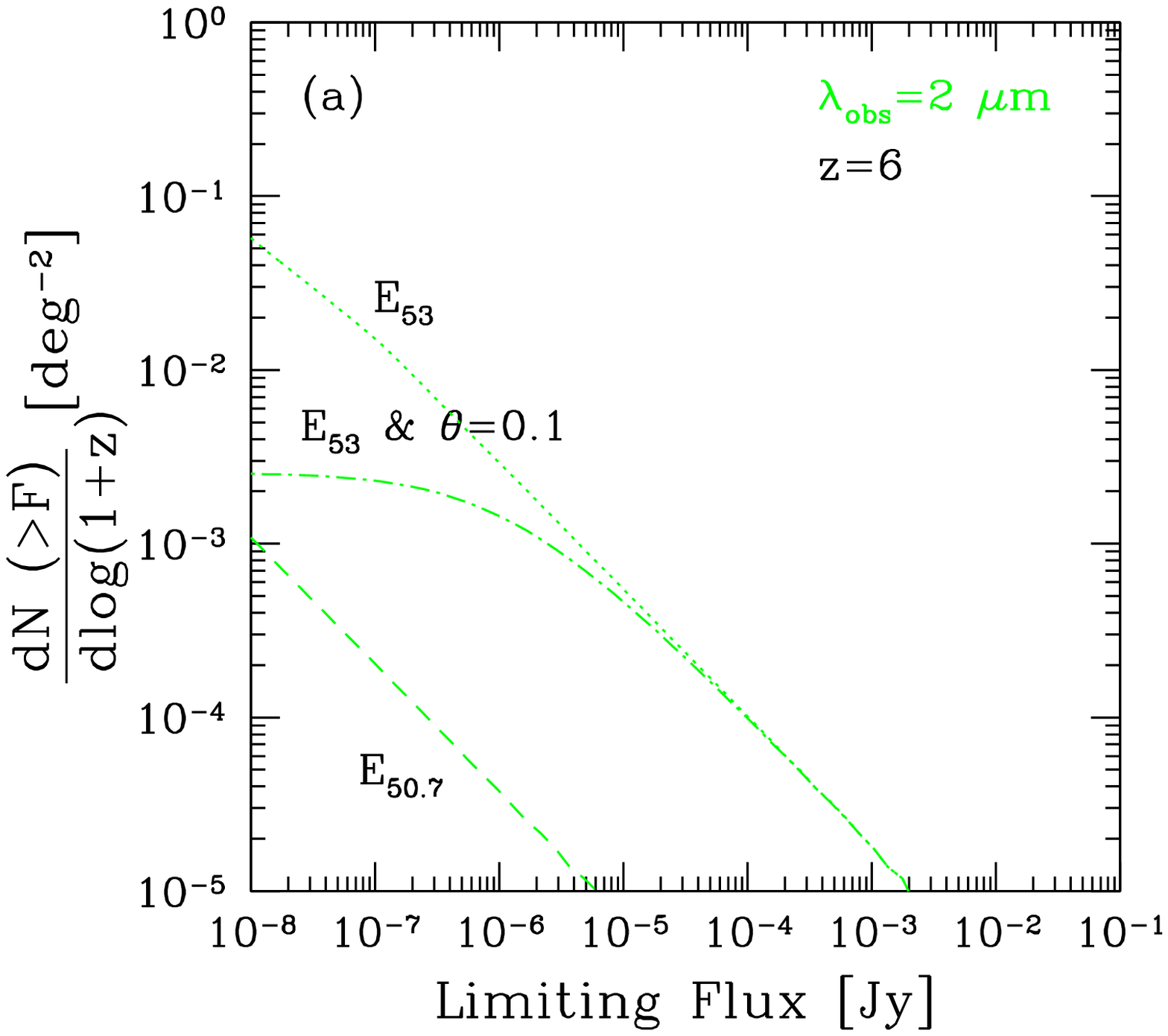}
\caption{\label{fig6}{(a) Contribution to the total number count of GRBs from a
redshift bin centered at $z=6$ for an observed wavelength of 2
$\mu$m. The different curves refer to different energy
outputs: $E=10^{53}$ erg (labeled $E_{53}$; dotted line), and $5 \times
10^{50}$ erg $=10^{50.7}$ erg (labeled $E_{50.7}$; dashed line).  Two of the
curves refer to a beamed explosion with an equivalent isotropic energy
output of $10^{53}~{\rm ergs}$ and a beaming angular radius of $\theta=0.1$
(dot-dashed line; see text for details). The actual energy output in this
case is $5\times 10^{50}$ erg.}}
\end{figure}
\setcounter{figure}{5}
\begin{figure}[t]
\plotone{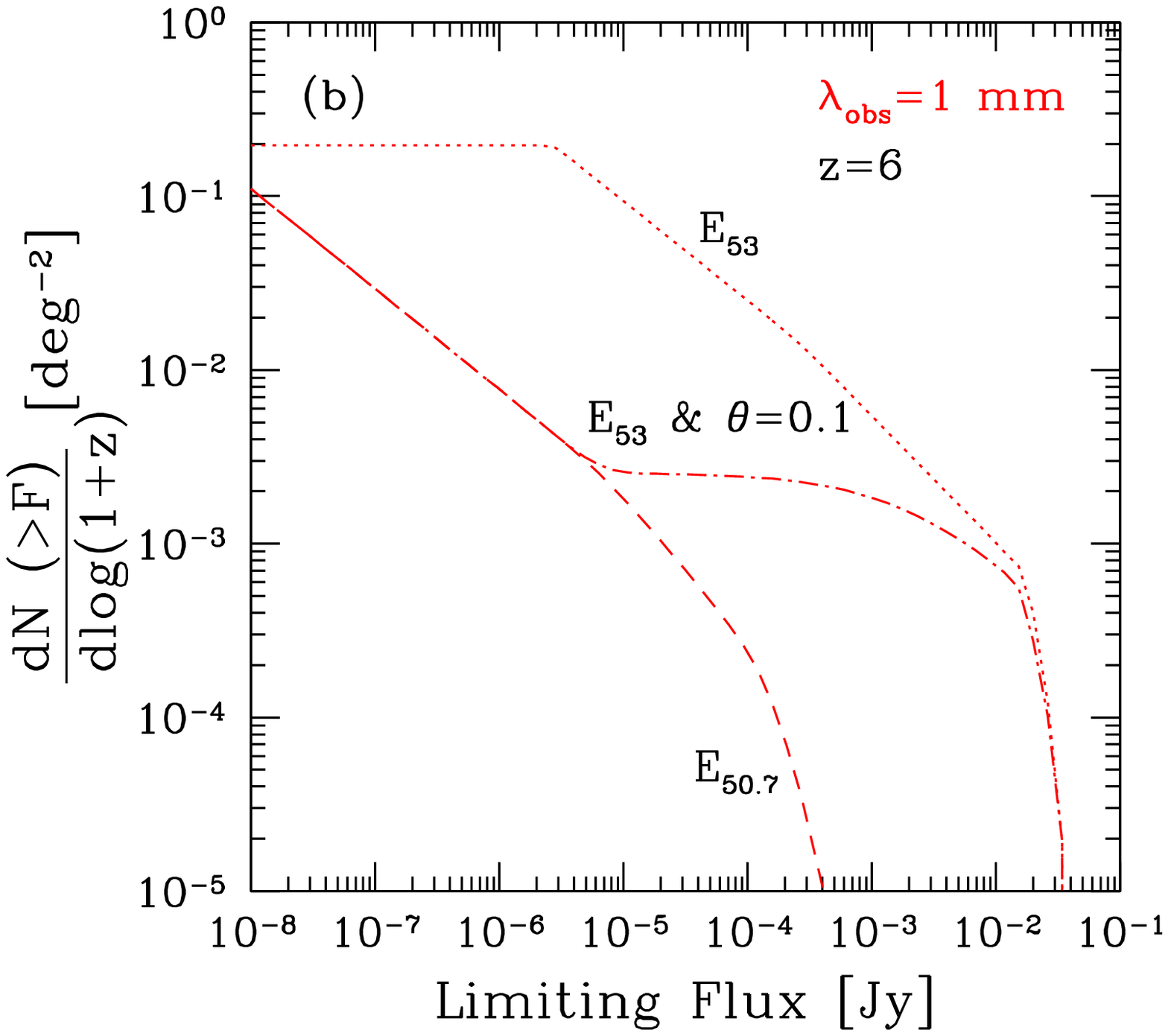}
\caption{{(b) As in Fig.~(6a), but for the reference observed wavelength of
1 mm.}}
\end{figure} 
\setcounter{figure}{5}
\begin{figure}[t]
\plotone{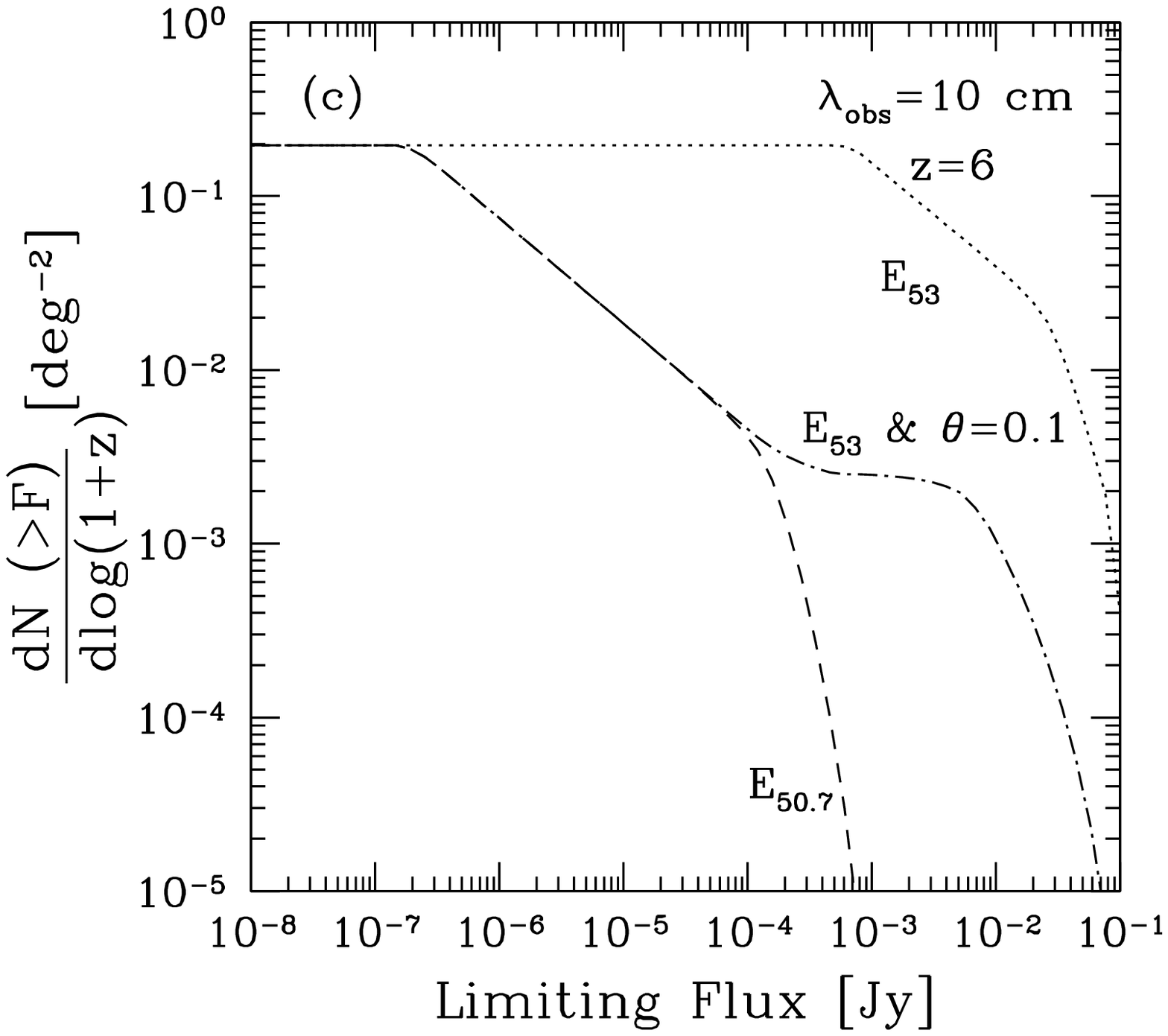}
\caption{{(c) As in Fig.~(6a), but for the reference observed wavelength of
10 cm.}}
\end{figure} 
\end{document}